\documentclass[11pt]{article}

\usepackage[T1]{fontenc}
\usepackage[margin=1in]{geometry}
\usepackage{lmodern}
\usepackage{amsmath}
\usepackage{amssymb}
\usepackage{graphicx}
\usepackage{booktabs}
\usepackage{tabularx}
\usepackage{caption}
\usepackage{microtype}
\usepackage{xcolor}
\usepackage{xspace}
\usepackage[numbers,sort&compress]{natbib}
\usepackage[hidelinks]{hyperref}
\hypersetup{
  pdftitle={Cross-Stack Validation of Language-Model Training},
  pdfauthor={Thang Tran, Lan Dang},
}

\graphicspath{{figures/}}
\newcommand{\NRecordsIn}{194{,}000\xspace}

\newcommand{\NTrain}{168{,}574\xspace}
\newcommand{\NEval}{3{,}440\xspace}
\newcommand{\PctAccounted}{100.0\%\xspace}

\newcommand{\NTrainable}{10{,}092{,}544\xspace}
\newcommand{\PctTrainable}{1.693\%\xspace}
\newcommand{\NSteps}{10{,}536\xspace}
\newcommand{\NEvalSamples}{512\xspace}

\newcommand{\PplStart}{12.57\xspace}
\newcommand{\PplEnd}{8.70\xspace}
\newcommand{\PplDrop}{31\%\xspace}
\newcommand{\LossStart}{2.5314\xspace}
\newcommand{\LossEnd}{2.1635\xspace}
\newcommand{\NEvals}{42\xspace}

\newcommand{\MeanAbsDev}{0.134\%\xspace}

\newcommand{\MaxDev}{0.316\%\xspace}
\newcommand{\StdDev}{0.138\%\xspace}
\newcommand{\SignChanges}{12\xspace}

\newcommand{\GradAgree}{0.1\%\xspace}
\newcommand{\NImpl}{four\xspace}
\newcommand{\NPerfImpl}{eight\xspace}

\newcommand{\SpreadPct}{0.15\%\xspace}

\newcommand{\GradSpread}{0.11\%\xspace}
\newcommand{\SDKSteps}{10{,}529\xspace}

\newcommand{\SDKPyMAD}{0.060\%\xspace}
\newcommand{\SDKZigMAD}{0.151\%\xspace}

\newcommand{\BaseLossSDK}{3.672389\xspace}
\newcommand{\BaseLossRef}{3.672389\xspace}
\newcommand{\HostToDevInt}{12.79\,GB/s\xspace}
\newcommand{\HostToDevExt}{2.73\,GB/s\xspace}

\newcommand{\VramInt}{199\,GB/s\xspace}
\newcommand{\VramExt}{330\,GB/s\xspace}

\newcommand{\PerfSteps}{500\xspace}
\newcommand{\PerfWarmup}{100\xspace}
\newcommand{\PerfMeasured}{400\xspace}

\newcommand{\PerfPtTok}{305\xspace}
\newcommand{\PerfFwTok}{457\xspace}

\newcommand{\GoLeakRatio}{2.6$\times$\xspace}

\newcommand{\GoLeakUpdates}{20\xspace}

\newcommand{\PerfBindSpread}{4.1\%\xspace}

\newcommand{\PerfProbeDrift}{2.0\%\xspace}

\newcommand{\FormatCost}{5.8\%\xspace}
\newcommand{\SmallMove}{0.0003\xspace}
\newcommand{\BigMove}{0.15\xspace}
\newcommand{\StopStep}{8{,}912\xspace}

\newcommand{\numbat}{\emph{numbat}\xspace}

\title{\textbf{Cross-Stack Validation of Language-Model Training:\\
A Clinical Fine-Tuning Case Study}}
\author{%
  Thang Tran\thanks{Corresponding author: \texttt{thang.tran@cloudkites.com}}\\
  \small CloudKites AI Lab\\
  \small New South Wales, Australia\\
  \small\texttt{thang.tran@cloudkites.com}
  \and
  Lan Dang\\
  \small Monash Business School, Monash University\\
  \small Victoria, Australia\\
  \small\texttt{LanHong.Dang@monash.edu}
}
\date{August 2026}

\begin{document}
\maketitle

\begin{abstract}
Neural network training suffers from an oracle problem. A run can converge normally, report a falling loss and yield a usable model even when the software beneath it computes something other than what was specified. Since almost all such work runs on one software stack, there is rarely anything independent to check it against. We ask whether independently implemented training stacks can act as differential oracles for a complete fine-tuning pipeline, rather than for the individual operators and inference paths targeted by earlier differential testing of deep-learning libraries.

The protocol we define works at the level of the learning trajectory. It comprises a shared specification, cross-check points spanning arithmetic, model loading, data rendering and the trajectory itself, and an explicit separation of three kinds of implementation independence: of the stack, of the training orchestration, and of the language runtime. We apply it to a realistic workload, a LoRA adaptation of Qwen3-0.6B over \NTrain{} clinical question-answer pairs, running the identical specification under PyTorch and under numbat, an independent framework written in Zig, driven natively and through its C interface from six programming languages.

The two stacks agree closely. Over \NEvals{} paired evaluations spanning a full epoch their held-out cross-entropy differs by \MeanAbsDev{} on average, and \NImpl{} implementations end the epoch within \SpreadPct{} of one another. The comparison also exposed 17 faults that single-implementation development had missed. Two of these matter for software engineering. The first is that the fault with the largest effect on the trained model lay outside the numerical kernels altogether: a mismatch in how clinical text was rendered moved held-out loss by \BigMove{}, roughly 500 times more than the arithmetic faults found alongside it. The second is that four faults were reachable only from a language whose memory model differs from the first two implementations, namely a scheduler that migrates work across threads, a collector that cannot observe device memory, and an ownership discipline requiring a primitive the interface did not provide. Implementation diversity, on this evidence, has several axes, and the runtime is one of them.
\end{abstract}

\section{Introduction}\label{sec:intro}

A trained model is the product of a long chain of arithmetic and data handling. When one link in that chain is wrong the result is rarely a crash. It is a slightly worse model, produced quietly, with a loss curve that still falls and outputs that still read fluently. The chain is hard to test for the same reason numerical software is generally hard to test: for most of the computation there is no independently known correct answer. Training compounds this in two ways. The computation is stateful, so an error at update $t$ propagates through every update that follows, and it is stochastic, so even two correct runs will not agree exactly.

Software engineering's established answer to a missing oracle is to obtain a second implementation and compare. The idea has been applied productively to deep-learning libraries, and \S\ref{sec:related} surveys that work. Almost all of it operates on individual API calls, on generated models, or on inference, so the unit under comparison is an operator or a forward pass. The closest work we are aware of that targets training itself compares distributed against non-distributed execution of the same library~\citep{wang2025d3}.

We ask instead what happens when the unit under comparison is a whole fine-tuning pipeline: template rendering, tokenisation, forward, loss, backward, optimizer, checkpoint and evaluation, executed by stacks that share no code. The learning trajectory then becomes the signal. Two stacks implementing the same specification should do more than produce similar final models. They should pass through the same sequence of held-out losses along the way, and should depart from that sequence together whenever the specification is misread.

\paragraph{Contributions.}
\begin{enumerate}
\item A trajectory-level differential validation protocol for end-to-end training (\S\ref{sec:design}). It fixes a shared specification, defines cross-check points at four levels from arithmetic through to the trajectory, and distinguishes three kinds of implementation independence that earlier work treats as one.
\item An empirical account of which observation exposes which fault (\S\ref{sec:faults}). Of the 17 faults, the one with the largest effect on the trained model was a data-rendering fault rather than a numerical one, and four were exposed only by runtime diversity. This gives evidence that independence of the language runtime is a distinct and productive axis, and connects the result to the classical warning that independently written versions can still fail together~\citep{knight1986nversion}.
\item A realistic case study (\S\ref{sec:workload}-\S\ref{sec:cost}) showing that the protocol is affordable: a full clinical adaptation reproduced across stacks, with the cost of running the second implementation measured rather than assumed.
\end{enumerate}

\paragraph{Scope.} We do not claim to introduce differential testing of machine-learning software, and \S\ref{sec:related} establishes that we do not. What we claim is a level of application and a decomposition of independence. On the clinical side we claim nothing about answer accuracy: the corpus supplies a realistic workload, and lower held-out loss on clinical text is not evidence that a model answers clinical questions correctly (\S\ref{sec:threats}). The adapted model is a research artifact and is not a medical device.

\section{Related work}\label{sec:related}

\paragraph{Differential testing of deep-learning libraries.} Using one implementation as the oracle for another is well established in this area. CRADLE~\citep{pham2019cradle} runs the same Keras model across backends and localises any divergence. LEMON~\citep{wang2020lemon} and Audee~\citep{guo2020audee} generate or mutate models to drive the comparison into rarely exercised code. FreeFuzz~\citep{wei2022freefuzz} mines argument values from open-source code to fuzz individual APIs at scale. A recent survey~\citep{survey2025dltesting} organises the field into model-level and API-level approaches, which fairly summarises where the unit of comparison has sat. Closest to our work is $D^3$~\citep{wang2025d3}, which does test training, comparing distributed against non-distributed execution using generated models. Its oracle is a second execution mode of one library; ours is a second stack, and our workload is a real adaptation rather than a generated model.

\paragraph{What goes wrong in these frameworks.} Empirical studies motivate the emphasis on silent failure.~\citep{chen2023dlbugs} classify 1,000 bugs across four frameworks, and~\citep{tambon2024silent} study bugs in Keras and TensorFlow that produce neither an error nor a crash, which is precisely the class a converging loss curve conceals. Our fault inventory (\S\ref{sec:faults}) is consistent with both, and adds a category their framework-level focus cannot reach: faults in the pipeline around the framework, such as how text is rendered before it is tokenised.

\paragraph{Design diversity.} The intellectual foundation is older. $N$-version programming~\citep{chen1978nversion} proposed independently implemented versions as a route to reliability, and~\citep{knight1986nversion} showed experimentally that the assumption of independent failure does not hold, since separately written versions make correlated mistakes. That warning is why \S\ref{sec:independence} declines to treat ``independent implementation'' as a single property. Our runtime-diversity result is a modern instance of the same lesson, though a more encouraging one: versions differing in a dimension the original pair shared do find faults the pair could not.

\paragraph{Reproducibility.} Work on reproducibility in machine learning is mostly concerned with re-running the same code and obtaining the same result. Our question is different and complementary: whether a separate implementation of the same specification produces the same learning behaviour. A field can have excellent reproducibility in the first sense while having no evidence at all in the second.

\paragraph{Gap.} Existing differential testing of machine-learning software
targets APIs, generated networks, inference, or alternative execution modes of a
single library. We study end-to-end training as a stateful stochastic
computation, use its learning trajectory as the differential signal, and ask
whether diversity in \emph{runtime semantics} exposes fault classes that stack
diversity alone does not.

\section{Study design}\label{sec:design}

\subsection{The shared specification}

All implementations consume one configuration file and one corpus, and Table~\ref{tab:recipe} states that configuration. Nothing determining how much arithmetic an update performs is left to an implementation's defaults: the adaptation method and rank, the optimizer and schedule, the batch composition, the sequence length, the precision and the evaluation protocol are all fixed. Where a framework's default disagrees with the specification, the specification takes precedence.

\begin{table}[t]
\caption{Adaptation configuration, identical across implementations.}
\label{tab:recipe}
\centering\small
\begin{tabularx}{\textwidth}{@{}l X@{}}
\toprule
\textbf{Adaptation} & Rank-stabilised LoRA~\citep{hu2021lora,kalajdzievski2023rslora},
rank 16, $\alpha=32$, dropout 0.05, on all seven projections of every layer
(query, key, value, output, gate, up, down); \NTrainable{} trainable
parameters (\PctTrainable{}). \\
\midrule
\textbf{Optimization} & AdamW~\citep{loshchilov2019adamw}, $\mathrm{lr}=10^{-4}$,
weight decay 0; cosine decay to zero after 3\% linear warmup; global-norm
gradient clipping at 1.0; effective batch 16 sequences (2 $\times$ 8
accumulation); 1 epoch, \NSteps{} updates. \\
\midrule
\textbf{Data} & Sequence length 512, truncation, no packing; loss on assistant
positions only; length-bucketed batching. \\
\midrule
\textbf{Precision} & 32-bit float throughout, in every implementation. \\
\midrule
\textbf{Evaluation} & Every 250 updates on a fixed \NEvalSamples{}-example
prefix of the held-out split --- a prefix, never a sample, so the curve moves
when the model does and not when the sample does. \\
\bottomrule
\end{tabularx}
\end{table}

\subsection{Three kinds of independence}\label{sec:independence}

``Independent implementation'' covers arrangements that differ considerably in how likely they are to fail together. Following the concern raised by~\citep{knight1986nversion}, we separate three.

Stack independence. PyTorch 2.11 with transformers and peft~\citep{paszke2019pytorch,wolf2020transformers,peft2022} against numbat~\citep{numbat2026yolo}, a self-contained toolkit of tensor operations, automatic differentiation and neural-network modules written from scratch in Zig with no third-party runtime dependency. Different kernels, different autograd, different language. This is the strongest form available to us, and the only one under which a shared numerical mistake is genuinely unlikely.

Orchestration independence. The native numbat trainer against training loops written independently above numbat's C application binary interface. The kernels are shared, so numerical agreement is expected and tells us little; what is under test is the loop, the adapter construction, the checkpointing strategy and the optimizer stepping, each written afresh. A fault in the interface, or in either loop, appears as disagreement.

\textbf{Runtime independence.} The same C interface driven from Zig, Python, Go,
Rust, TypeScript and C. The arithmetic is identical by construction. What
differs is the scheduler, the memory model, the ownership discipline and the
foreign-function semantics. \S\ref{sec:faults} reports that this axis, which
contributes nothing numerically, exposed four faults the other two could not.

Reporting these separately matters because a reader is otherwise invited to
count implementations and treat the count as strength. Six bindings over one
kernel set are not six independent tests of the arithmetic. They are one test of
the arithmetic and six tests of everything else.

\subsection{Cross-check points}\label{sec:checkpoints}

Agreement is assessed at four levels, ordered so that a failure at one level
explains failures above it.

\begin{description}
\item[L1, arithmetic.] Per-operator numerical checks and, before training, the
  untrained model's held-out loss. Both implementations must report the same
  loss for the same weights on the same data before either takes a step. This
  isolates the forward path from everything the training loop does.
\item[L2, model loading.] Parameter-by-parameter comparison after each
  implementation has loaded the released checkpoint. A loading fault is
  otherwise indistinguishable from a training fault.
\item[L3, data rendering.] The exact token sequence each implementation
  produces for the same record, including the chat template, the position of
  the supervision boundary, and truncation. \S\ref{sec:faults} explains why this
  level earns its place.
\item[L4, trajectory.] Held-out cross-entropy at each of \NEvals{} paired
  evaluation points across the epoch, plus gradient-norm distributions. This is
  the level at which faults invisible to L1--L3 appear.
\end{description}

\paragraph{What agreement at L4 does and does not establish.} Equal held-out
loss does not imply equal functions: two different models can achieve the same
average loss on the same data. The inference we draw is weaker and, we think,
still worth drawing. Agreement of independently computed observables at many
points along the trajectory is evidence that the implementations realise
materially equivalent training behaviour under the tested configuration. It is
evidence proportional to the number and diversity of the observables, which is
why \S\ref{sec:threats} lists richer observables as the most valuable extension
to this work.

\subsection{Apparatus}\label{sec:apparatus}

A single workstation (Intel Core i7-13800H, 6 cores / 12 threads, 31.7\,GB) with
two accelerators: an NVIDIA RTX 2000 Ada Laptop (8\,GB) internally connected, and
an NVIDIA GeForce RTX 3060 (12\,GB) in an external enclosure. Driver 595.79,
CUDA 13.2.

\paragraph{Every measurement was taken on Windows.} Table~\ref{tab:software}
records the environment. We state it because deep-learning work is very largely
developed and reported on Linux, and a reader is entitled to assume Linux unless
told otherwise. Nothing here required a Linux host, a container, or the Windows
Subsystem for Linux.

\begin{table}[t]
\caption{The software environment, recorded because it is unusual for this kind
of work and because two of the study's findings depend on it. Every leg ran on
this one installation.}
\label{tab:software}
\centering\small
\begin{tabular}{@{}ll@{}}
\toprule
Component & Version \\
\midrule
Operating system & Windows 11 Pro 25H2 (build 26200.8655) \\
GPU driver model & WDDM (not TCC; unavailable on a laptop part) \\
NVIDIA driver / CUDA & 595.79 / 13.2, cuDNN 9.20 \\
\midrule
\numbat{} (framework and C ABI) & Zig 0.17.0-dev.1564+97ced1272 \\
Reference stack & PyTorch 2.11.0+cu128, \texttt{transformers} 5.15.0, \texttt{peft} 0.20.0 \\
\midrule
SDK leg: Python & CPython 3.11 (\texttt{ctypes}; no third-party package) \\
SDK leg: Go & Go 1.26.5 (\texttt{cgo}) \\
SDK leg: Rust & rustc 1.97.1, \texttt{x86\_64-pc-windows-gnu} \\
SDK leg: TypeScript & Node.js 24.18.0 (N-API addon) \\
SDK leg: C & compiled with \texttt{zig cc} (C11) \\
\bottomrule
\end{tabular}
\end{table}

One property of the platform shapes a result later in the paper. Windows drives
consumer and mobile GPUs through WDDM, in which the operating system's video
memory manager owns device allocations and may page them to host memory under
pressure; the alternative model that does not (TCC) is unavailable on a laptop
part. An allocation that exceeds what the device can hold therefore degrades
rather than fails. Fault~17 in \S\ref{sec:faults} presents exactly that way: a
leg whose evaluation phase raised the allocator's high-water mark ran every
subsequent update roughly seven times slower, with losses unchanged and no error
raised. On a platform that returns an out-of-memory error the same fault would
have announced itself at once.

\paragraph{The two accelerators are not interchangeable.}
Table~\ref{tab:apparatus} measures them. On the host-to-device path the internal
card reaches \HostToDevInt{} against the external card's \HostToDevExt{}; on
device memory bandwidth the ordering reverses, \VramExt{} against \VramInt{}.
Neither card is simply faster.

\begin{table}[t]
\caption{The two accelerators, measured rather than quoted from specification. Transfers are 256\,MB through pinned host memory, timed with device events, best of 20 after warmup, with both cards verified idle. The device-copy column is a read-plus-write of a large buffer and is a proxy for memory bandwidth. Note the last two rows move in OPPOSITE directions: there is no ordering of these two cards.}
\label{tab:apparatus}
\centering\small
\begin{tabular}{@{}lrr@{}}
\toprule
& Internal & External \\
\midrule
Device & RTX 2000 Ada & RTX 3060 \\
Memory & 8\,GB & 12\,GB \\
PCIe link, negotiated & gen\,4 $\times$8 & gen\,4 $\times$4 \\
\midrule
Host $\rightarrow$ device & 12.79\,GB/s & 2.73\,GB/s \\
Device $\rightarrow$ host & 13.17\,GB/s & 3.12\,GB/s \\
\quad as \% of the negotiated link & 81\% & 35\% \\
Device copy (memory bandwidth) & 199\,GB/s & 330\,GB/s \\
\bottomrule
\end{tabular}
\end{table}

The trajectory runs of \S\ref{sec:agreement} executed concurrently, one on each
device, because running them in series would have cost some eighty hours. That
is sound for comparing loss at a \emph{step} and useless for comparing loss at a
\emph{time}. Accordingly those runs support no timing claim whatever; the
separately controlled measurement in \S\ref{sec:cost} does, on one card with one
implementation at a time.

\section{The workload}\label{sec:workload}

The protocol needs a workload that is realistic rather than synthetic: long
sequences, a non-trivial template, a real corpus, an adaptation method in
current use. A clinical instruction-tuning task supplies all four.

\subsection{Corpus}

\NRecordsIn{} rows were assembled from three publicly licensed sources
(Table~\ref{tab:corpora}), converted to a single instruction format, and
deduplicated, yielding \NTrain{} training and \NEval{} held-out examples; every
input row is accounted for as converted, duplicate or rejected
(\PctAccounted{}). In each source the training target is the written
\emph{explanation} rather than the letter of the correct option, since training
on ``(c)'' teaches a model to guess letters.

\begin{table}[t]
\caption{Source corpora. In every case the training target is the written
\emph{explanation}, not the letter of the correct option.}
\label{tab:corpora}
\centering\small
\begin{tabularx}{\textwidth}{@{}l X r l@{}}
\toprule
Source & Origin & Rows & Licence \\
\midrule
PubMedQA (\texttt{pqa\_labeled})~\citep{jin2019pubmedqa} &
  Questions derived from PubMed abstract titles with expert long-form answers &
  1{,}000 & MIT \\
MedMCQA~\citep{pal2022medmcqa} &
  Indian postgraduate medical entrance examinations (AIIMS, NEET-PG) &
  182{,}822 & Apache-2.0 \\
MedQA-USMLE (4-option)~\citep{jin2020medqa} &
  United States Medical Licensing Examination-style questions &
  10{,}178 & CC-BY-4.0 \\
\bottomrule
\end{tabularx}
\end{table}

This is examination material, not clinical practice. It contains no patient
data, and the distinction matters for interpreting any result on it: questions
are self-contained, well-posed and answerable from the text, which real
presentations are not.

\subsection{Model and adaptation}

Qwen3-0.6B~\citep{qwen3}, 596M parameters, adapted with low-rank
adaptation~\citep{hu2021lora}: the released weights are frozen and a small
number of trainable matrices is added beside them, so the adaptation is a 40\,MB
file and the base model is unchanged. Table~\ref{tab:recipe} gives the
configuration.

Two properties make this a good subject rather than merely a convenient one. The
model is small enough that a full epoch is affordable on one consumer card,
which is what makes running it under many implementations possible at all. And
the template is fussy: Qwen3 opens its final assistant turn with an empty
reasoning block, a detail that is easy to omit and that \S\ref{sec:faults} shows
is worth more than the arithmetic.

\section{Results: trajectory agreement}\label{sec:agreement}

\subsection{The adaptation}

Over one epoch (\NSteps{} updates) held-out perplexity fell from \PplStart{} to
\PplEnd{}, a \PplDrop{} reduction, and held-out cross-entropy from \LossStart{}
to \LossEnd{}. Every one of the \NEvals{} evaluations improved on the one
before. This establishes that the workload is a real learning task and not a
degenerate one; it establishes nothing about clinical correctness
(\S\ref{sec:threats}).

\subsection{Two stacks}

At L1, both implementations report held-out cross-entropy \BaseLossRef{} for the
untrained model --- agreement to six decimal places before either takes a step,
which places the forward path and the data pipeline beyond suspicion for what
follows.

\begin{figure}[t]
  \centering
  \includegraphics[width=\textwidth]{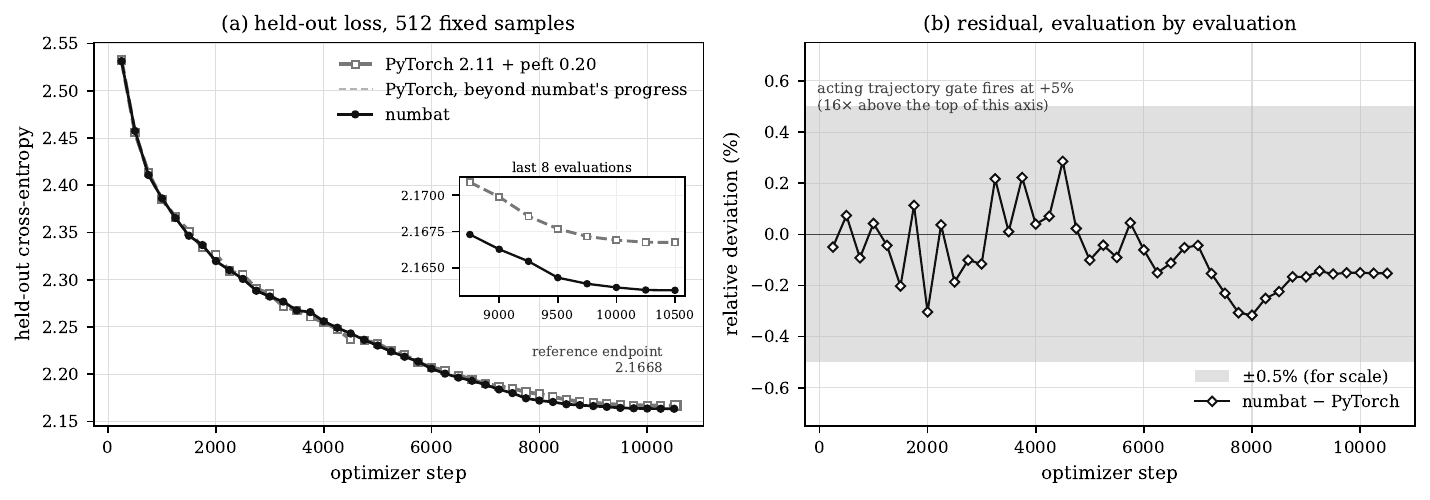}
  \caption{Held-out cross-entropy at each paired evaluation, both stacks, and
  the residual between them.}
  \label{fig:parity}
\end{figure}

Across the epoch (Figure~\ref{fig:parity}, Table~\ref{tab:agreement}), the two
stacks' held-out cross-entropy differs by \MeanAbsDev{} on average, with a
largest single difference of \MaxDev{} and a standard deviation of \StdDev{}.
The signed residual changes sign \SignChanges{} times, which is what an
unbiased difference looks like; a systematic numerical difference would show as
a residual of consistent sign.

\begin{table}[t]
\caption{Agreement between \numbat{} and the PyTorch reference, over every evaluation at which both reported a held-out loss on the identical fixed subset.}
\label{tab:agreement}
\centering\small
\begin{tabular}{@{}lr@{}}
\toprule
Statistic & Value \\
\midrule
Evaluations compared & 42 (steps 250--10{,}500) \\
Mean signed difference (bias) & $-$0.078\% \\
Mean absolute difference & 0.134\% \\
Dispersion of the difference & 0.138\% \\
Largest single difference & 0.316\% \\
Sign changes & 12 of 41 transitions \\
\midrule
\emph{numbat} held-out loss, first $\rightarrow$ final & 2.5314 $\rightarrow$ 2.1635 \\
PyTorch held-out loss, first $\rightarrow$ final & 2.5327 $\rightarrow$ 2.1668 \\
\bottomrule
\end{tabular}
\end{table}

Gradient norms are an internal quantity neither implementation reports as its
result, which makes them a useful second observable: an implementation could in
principle match on loss while taking different steps.
Table~\ref{tab:gradnorm} shows median gradient norms agreeing to \GradAgree{}.
The maxima differ considerably more, which we read as the tail being dominated
by individual batches whose composition differs between implementations rather
than as disagreement about the gradient.

\begin{table}[t]
\caption{Pre-clip gradient magnitude after warmup (step $\geq$ 500) --- an internal quantity no implementation reports as its result, and therefore an independent check on the agreement of Table~\ref{tab:fourway}.}
\label{tab:gradnorm}
\centering\small
\begin{tabular}{@{}lrrrr@{}}
\toprule
Implementation & Median & p99 & Max & Steps above clip \\
\midrule
PyTorch + \texttt{peft} (10{,}031 steps) & 3.238 & 5.984 & 12.34 & 100\% \\
\emph{numbat} framework (10{,}037 steps) & 3.240 & 6.063 & 27.55 & 100\% \\
\emph{numbat} SDK (Python) (10{,}030 steps) & 3.241 & 5.952 & 13.91 & 100\% \\
\emph{numbat} SDK (Zig) (10{,}030 steps) & 3.241 & 5.927 & 13.70 & 100\% \\
\bottomrule
\end{tabular}
\end{table}

\begin{figure}[t]
  \centering
  \includegraphics[width=\textwidth]{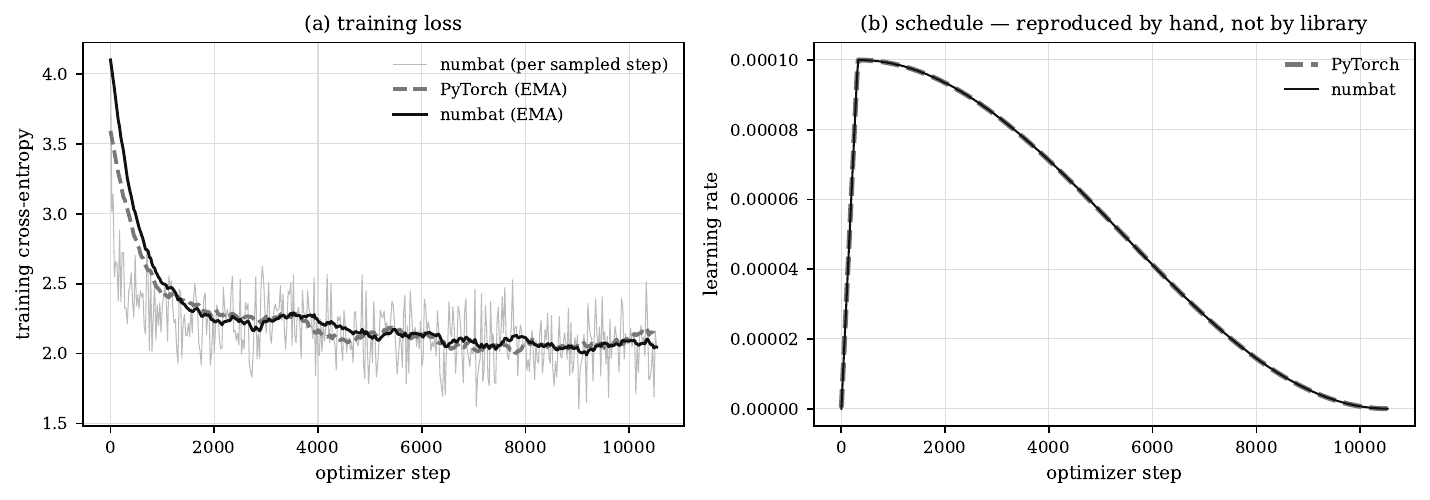}
  \caption{Training loss for both stacks over the epoch.}
  \label{fig:train}
\end{figure}

\subsection{Orchestration independence}

\begin{figure}[t]
  \centering
  \includegraphics[width=\textwidth]{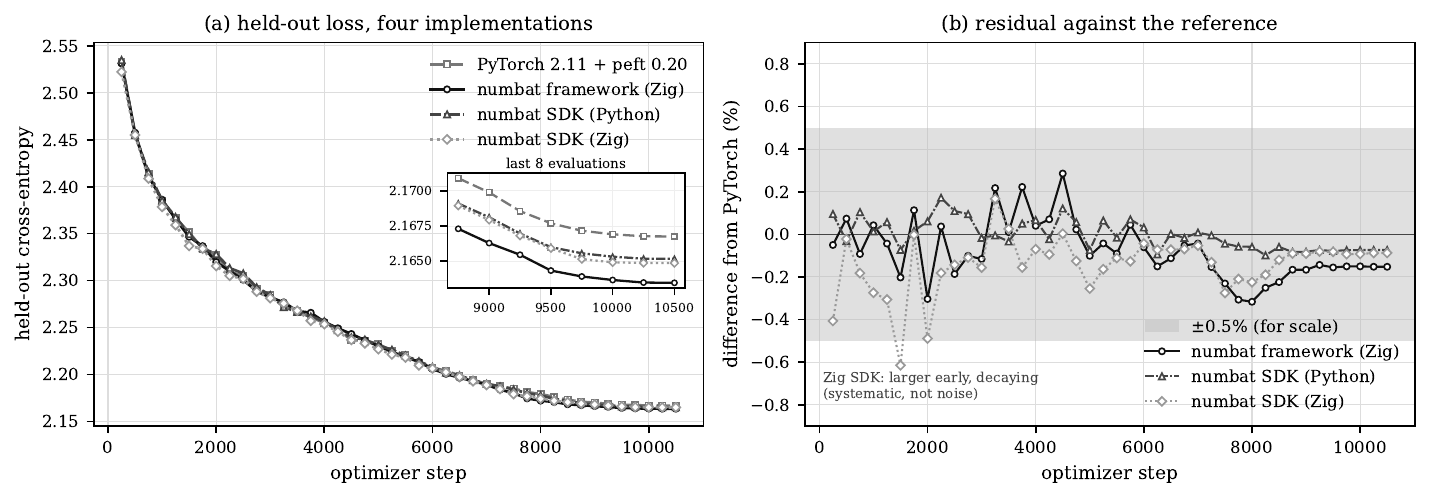}
  \caption{All four implementations over the same schedule.}
  \label{fig:fourway}
\end{figure}

Two further implementations drove \numbat{} through its C interface with
independently written training loops, one in Python and one in Zig, over
\SDKSteps{} updates. Both reproduced the untrained model's held-out loss to six
decimal places (\BaseLossSDK{}) and then tracked the reference: mean absolute
deviation \SDKZigMAD{} (Zig) and \SDKPyMAD{} (Python).
Table~\ref{tab:fourway} and Figure~\ref{fig:fourway} give the four-way
comparison; the implementations end the epoch within \SpreadPct{} of one another
and agree to \GradSpread{} on gradient magnitude.

\begin{table}[t]
\caption{The four implementations after one complete pass over the corpus. Each shares no training code with the others; all read the same configuration file and the same corpus, and are scored on the identical held-out subset. Differences are against the PyTorch reference, over the 42 evaluations each implementation shares with it.}
\label{tab:fourway}
\centering\small
\begin{tabular}{@{}llrrrr@{}}
\toprule
Implementation & Stack & Final loss & Mean abs.\ diff.\ & Largest diff.\ & Grad.\ median \\
\midrule
PyTorch + \texttt{peft} & Python & 2.1668 & \emph{reference} & \emph{reference} & 3.238 \\
\numbat{} framework & Zig & 2.1635 & 0.134\% & 0.316\% & 3.240 \\
\numbat{} SDK & Python / C ABI & 2.1652 & 0.060\% & 0.172\% & 3.241 \\
\numbat{} SDK & Zig / C ABI & 2.1649 & 0.151\% & 0.614\% & 3.241 \\
\midrule
\multicolumn{2}{@{}l}{Spread across all 4} & 2.1635--2.1668 & \multicolumn{3}{r@{}}{0.15\% of the lowest} \\
\bottomrule
\end{tabular}
\end{table}

As \S\ref{sec:independence} warned, this agreement is weaker evidence than the
two-stack comparison, because these implementations share kernels. Its value is
in what it does test: the interface, and three separately written training
loops.

\section{Results: which observation exposed which fault}\label{sec:faults}

Seventeen faults were identified during the study. They were not all silent, and
calling them so would overstate the case: five produced a hard failure
(an out-of-memory condition, a kernel that would not launch, autograd refusing a
gradient edge). What they have in common is narrower and sufficient --- each
escaped single-implementation development, and each was found by comparing
implementations rather than by testing one.

\begin{table*}[t]
\caption{Which observation exposed which fault. Reconstructed from the
development record: for each fault we record the signal that \emph{did} expose
it and, where the record is unambiguous, signals that were active and did not.
\checkmark{} = exposed the fault; $\circ$ = active but did not expose it;
\textendash{} = not applicable or not active at the time. A controlled
re-injection study (\S\ref{sec:threats}) would replace the reconstruction with a
measurement; the asymmetry visible here is nonetheless the paper's central
empirical claim, and it is that the rightmost two columns cover faults no other
column does.}
\label{tab:faultmatrix}
\centering\footnotesize
\begin{tabular}{@{}llccccc@{}}
\toprule
& & \multicolumn{5}{c}{Observation} \\
\cmidrule(l){3-7}
\# & Fault & Unit / op & Initial & Data-render & Trajectory & Runtime \\
& & test & forward & cross-check & agreement & diversity \\
\midrule
1  & Chat template not the model's        & $\circ$ & $\circ$ & \checkmark & \checkmark & \textendash \\
2  & Adapter placement (shared projection) & $\circ$ & $\circ$ & \textendash & \checkmark & \textendash \\
3  & Transpose-matmul backward on views   & \checkmark & $\circ$ & \textendash & \checkmark & \textendash \\
4  & Dropout mask drawn on the host       & $\circ$ & $\circ$ & \textendash & $\circ$ & \textendash \\
5  & Attention kernel shared-memory opt-in & \checkmark & \checkmark & \textendash & \textendash & \textendash \\
6  & Memory options silently discarded    & $\circ$ & \checkmark & \textendash & \textendash & \textendash \\
7  & Precision option parsed and ignored  & $\circ$ & $\circ$ & \textendash & $\circ$ & \textendash \\
8--11 & Configuration and reporting       & $\circ$ & $\circ$ & \textendash & $\circ$ & \textendash \\
\midrule
12 & Corpus read not bounded by the work  & $\circ$ & $\circ$ & \checkmark & $\circ$ & \textendash \\
13 & Checkpoint re-read per tensor        & $\circ$ & $\circ$ & \textendash & $\circ$ & \textendash \\
\midrule
14 & Per-thread state vs.\ green threads  & $\circ$ & $\circ$ & \textendash & $\circ$ & \checkmark \\
15 & Collector cannot see device memory   & $\circ$ & $\circ$ & \textendash & $\circ$ & \checkmark \\
16 & No second owning handle in the ABI   & $\circ$ & $\circ$ & \textendash & \textendash & \checkmark \\
17 & Evaluation ran without the adapter   & $\circ$ & $\circ$ & \textendash & \checkmark & \checkmark \\
\bottomrule
\end{tabular}
\end{table*}

Table~\ref{tab:faultmatrix} records which observation exposed each. It is a
reconstruction from the development record rather than a controlled experiment,
a limitation \S\ref{sec:threats} takes seriously. Three patterns in it are worth
drawing out.

\paragraph{The largest fault was not a numerical one.} Faults 2 and 3 --- an
adapter placed on a shared projection, and a transpose-matmul reading
non-contiguous operands in its backward pass --- are exactly the kind of
arithmetic error differential testing is built to find, and fixing them moved
held-out loss by \SmallMove{}. Fault~1 was that the training text was rendered
under a generic chat layout rather than the one the base model was taught.
Fixing that moved held-out loss by \BigMove{}, roughly 500 times more, and it
had put the run \FormatCost{} above the reference while every training metric
looked healthy. The dominant correctness risk in this pipeline lived at a
semantic boundary, not a numerical one. Framework-level differential testing
would not have looked there, because the fault is not in the framework.

\paragraph{Runtime diversity found what stack diversity could not.} Faults
14--17 were reachable only from a language whose memory model differs from the
first two implementations.

\begin{itemize}
\item \textbf{Per-thread state against a work-stealing scheduler.} \numbat{}
  keeps its no-grad flag, the interface's tensor scope stack and the random
  number generator in thread-local storage, which is correct given that a device
  context belongs to a thread. Go's scheduler moves a goroutine between
  operating-system threads at foreign-call boundaries, so a forward pass began
  on one thread with the graph enabled and continued on another where it was
  not. It presented as autograd refusing a gradient edge \emph{at a different
  layer on every run}. Neither Python (one thread under the interpreter lock)
  nor Zig (no scheduler) can produce it.
\item \textbf{A collector cannot see device memory.} A tensor handle in Go is
  eight bytes of collected heap in front of tens of megabytes of device memory.
  The collector sees no pressure, never runs, and the finalisers that would
  release those tensors never fire; the caching allocator then trims and
  synchronises on every allocation. Measured: \GoLeakRatio{} slower by update
  \GoLeakUpdates{} with identical losses, and a device-memory trace \emph{flat}
  at the ceiling, which reads as saturation rather than as a leak. Reference
  counting (Python) and manual memory (Zig, C) make the question invisible.
\item \textbf{An owning handle cannot be in two places.} Rust's tensor owns its
  handle and releases on drop, so the same adapter parameter cannot be given to
  an optimizer that takes ownership and also kept by the adapter hook. The fix
  was a primitive the interface did not have --- a second owning handle over one
  tensor --- which the C implementation then used as well.
\item \textbf{Evaluation ran without the adapter.} Held-out loss sat at the
  untrained value at every checkpoint. A flat held-out curve reads as a training
  failure; here it was an evaluation fault, and fixing it exposed the memory
  behaviour described in \S\ref{sec:apparatus}.
\end{itemize}

\paragraph{The levels are not redundant.} No single column of Table~\ref{tab:faultmatrix} covers the whole table. Unit tests caught the two faults that were genuinely local to an operator. The initial forward comparison caught configuration faults that would otherwise have been attributed to training. Data-rendering cross-checks caught the most consequential fault in the study. Trajectory agreement caught the faults that stay invisible until state accumulates. Runtime diversity caught a class that the other four cannot reach. A protocol stopping at any one of these would have shipped the remainder.

\section{Feasibility}\label{sec:cost}

Independent validation is only useful if running the second implementation is affordable, so we measured that rather than assuming it. The measurement is a study separate from \S\ref{sec:agreement} and designed for the purpose: \NPerfImpl{} implementations executed the configuration of Table~\ref{tab:recipe} for \PerfSteps{} updates on the same accelerator, one at a time, on an otherwise idle machine, with the device verified idle before and after each. The first \PerfWarmup{} updates are discarded as warmup, leaving \PerfMeasured{} measured.

\begin{table}[t]
\caption{The controlled study: 500 updates of the identical recipe, one implementation at a time, on the internally connected card, with the device verified idle before and after each leg. Per-update cost is the median over steps 101--500; the first 100 are discarded as warmup. Setup is model load plus corpus tokenisation, reported separately because being slow to tokenise is a different defect from being slow to train.}
\label{tab:perfstudy}
\centering\small
\begin{tabular}{@{}llrrrrr@{}}
\toprule
Implementation & Stack & s/update & tok/s & vs.\ best & Setup (s) & Peak MiB \\
\midrule
PyTorch 2.11 + peft & Python & 4.600 & 305 & 1.12$\times$ & 120 & 7947 \\
\numbat{} SDK (Zig) & Zig / C ABI & 4.354 & 437 & 1.06$\times$ & 10 & 7481 \\
\numbat{} framework & Zig & 4.100 & 457 & 1.00$\times$ & 20 & 7461 \\
\numbat{} SDK (Python) & Python / C ABI & 4.327 & 438 & 1.06$\times$ & 10 & 7461 \\
\numbat{} SDK (Go) & Go / C ABI & 4.427 & 432 & 1.08$\times$ & 11 & 7552 \\
\numbat{} SDK (Rust) & Rust / C ABI & 4.181 & 450 & 1.02$\times$ & 9 & 7552 \\
\numbat{} SDK (TypeScript) & Node / C ABI & 4.345 & 431 & 1.06$\times$ & 12 & 7552 \\
\numbat{} SDK (C) & C / C ABI & 4.290 & 440 & 1.05$\times$ & 12 & 7544 \\
\bottomrule
\end{tabular}
\end{table}

Every numbat implementation completed the study faster than the reference and in less device memory: the framework at \PerfFwTok{} trained tokens per second against \PerfPtTok{}, with the six bindings within \PerfBindSpread{} of one another. Setup differs more than training does, and we report it separately, since being slow to tokenise is a different defect from being slow to train.

We deliberately do not develop this into a performance claim. One run per implementation, on a machine whose clock behaviour varies between sessions by more than the differences being reported, cannot support one; \S\ref{sec:threats} gives the detail. What the measurement does support is the feasibility statement the paper needs: running a second, independent implementation of this workload costs about as much as running the first, so the validation protocol is not priced out of use.

\section{Discussion}\label{sec:discussion}

\paragraph{Independence is not one property.} Our most transferable finding is that the fault classes an independent implementation can expose depend on how it is independent. Stack independence covers the arithmetic. Orchestration independence covers the training loop and its interface to the kernels. Runtime independence covers scheduling, lifetime and ownership, and covers nothing numerical at all. A second implementation written in a language that manages memory the same way as the first is not fully independent, and the faults it cannot find are precisely the quiet ones. This is the correlated-failure result of~\citep{knight1986nversion} arriving in a new setting, and it suggests choosing a second implementation for the dimensions in which it differs rather than simply for being second.

\paragraph{Semantic boundaries deserve the scrutiny numerical ones get.} The 500-fold difference between the effect of the data-rendering fault and that of the arithmetic faults is a single observation and should not be over-read. It does, however, point somewhere the literature surveyed in \S\ref{sec:related} does not look. Differential testing of frameworks tests frameworks, whereas the template deciding what text the model is trained on sits above the framework, is written per project, and is covered by none of it.

\paragraph{For practitioners.} The cheapest useful step is L1: run the untrained model through both implementations and compare held-out loss before training anything. It costs minutes, needs no second training run, and would have caught several of the faults reported here. The next cheapest is L3, comparing rendered token sequences for a handful of records, which is where the most expensive fault in this study was hiding.

\section{Threats to validity}\label{sec:threats}

\paragraph{The fault matrix is reconstructed, not measured.} Table~\ref{tab:faultmatrix} records what did expose each fault during development, not what would expose it under controlled re-injection. Faults were found and fixed in a particular order, and a signal that fired second might well have fired first had that order differed. A re-injection study, restoring each fault individually and running every observation against it, would turn the table into a measurement. We regard it as the single most valuable extension to this work.

\paragraph{One scalar carries most of the trajectory evidence.} Held-out cross-entropy together with gradient norm is a thin comparison vector for a claim about equivalent training behaviour. Richer observables, such as prediction distributions on a fixed probe set, parameter-update summaries and activation statistics at fixed positions, would allow the claim of \S\ref{sec:checkpoints} to be stated far more strongly. We regard their absence as the principal methodological weakness of this study.

\paragraph{No stochastic baseline.} We interpret residuals such as \MeanAbsDev{} as small, but we have not established what run-to-run variation looks like within a single implementation across different seeds. Without that envelope, ``small'' is a judgement rather than a measurement: the right comparison is between cross-stack distance and same-stack replicate distance, and we have only the former. For the same reason, we used one seed throughout.

\paragraph{One trajectory run was interrupted.} The numbat epoch was stopped at update \StopStep{} of \NSteps{} by a supervisory rule that fired on a single noisy batch, and was resumed from its last checkpoint. Because optimizer moments were not checkpointed, the resumed curve carries one discontinuity. We report this rather than presenting the trajectory as unbroken; a clean uninterrupted rerun would remove the caveat.

\paragraph{The timing study supports feasibility only.} There is one run per implementation, and the implementations ran on different days on a host whose clock governor is not stable across sessions: an earlier attempt at one leg, on bit-identical batches, ran five to eight times slower than the leg reported here. Position in the sequence is matched, and the reference was re-measured before and after the other legs (drift \PerfProbeDrift{}), but these are mitigations rather than controls. Repeated interleaved runs with confidence intervals would be needed for any comparative performance claim, and we make none.

\paragraph{Scale and configuration.} The study covers one 596M-parameter model, one adaptation method, f32 precision and one laptop-class card. Nothing here speaks to larger models, reduced precision, multiple devices, or the distributed regimes where much framework engineering is concentrated. Absolute memory figures are additionally properties of the platform (\S\ref{sec:apparatus}).

\paragraph{Clinical.} No downstream accuracy is measured. Lower held-out loss means the model predicts clinical answer text better, not that its answers are right, and a model can improve on the former while remaining unsafe on the latter. No safety evaluation was performed. The artifact is for research use and is not a medical device.

\section{Conclusion}\label{sec:conclusion}

Training software can implement the wrong computation and still converge. We have shown that an independently implemented training stack can act as a differential oracle for an entire fine-tuning pipeline, that the learning trajectory is a usable signal for the comparison, and that the comparison is affordable enough to be worth running.

Applying the protocol to a clinical adaptation of Qwen3-0.6B, two stacks sharing
no training code agreed to \MeanAbsDev{} in held-out cross-entropy across
\NEvals{} paired evaluations over a full epoch, and \NImpl{} implementations
finished within \SpreadPct{} of one another. Seventeen faults surfaced that
single-implementation development had not.

Two results are worth carrying forward, and both concern where faults hide. The most damaging fault was not in the arithmetic but in how clinical text was rendered before the model saw it, and it cost roughly 500 times more held-out loss than the numerical faults found beside it. Four further faults were reachable only from a language whose scheduler, collector or ownership model differed from the first two implementations: faults that two independent stacks, both managing memory in the same way, could not have found between them.

This suggests a practical rule for anyone building a second implementation as a check on the first. Diversity of implementation is not a count. It has dimensions, namely stack, orchestration and runtime, and a second version is worth most in those dimensions where it differs.

\section*{Data and code availability}

Measurement data supporting every figure and table, together with the scripts that generate them, are available from the authors. The adapted model is a research artifact and is not a medical device.

\bibliographystyle{plainnat}
\bibliography{references}

\end{document}